\documentclass[aps,prb,twocolumn,superscriptaddress,longbibliography,showpacs]{revtex4-1}

\usepackage{graphicx}
\usepackage{graphics}
\usepackage{dsfont}
\usepackage{bm}
\usepackage{amsmath}

\newcommand{\eop}{\hat{c}}
\newcommand{\hop}{\hat{d}}
\newcommand{\Hop}{\hat{H}}
\newcommand{\eopd}{\hat{c}^{\dagger}}
\newcommand{\hopd}{\hat{d}^{\dagger}}
\newcommand{\kvec}{\textbf{k}}
\newcommand{\kdashvec}{\textbf{k}'}
\newcommand{\qvec}{\textbf{q}}
\newcommand{\Efield}{\textbf{E}}
\newcommand{\Vkkd}{V_{\left|\kvec - \kvec'\right|}}

\DeclareMathOperator{\sech}{sech}

\newcommand*\dif{\mathop{}\!\mathrm{d}}

\begin{document}

\title{Dynamical calculation of third harmonic generation in a semiconductor quantum well}

\author{Stefano Guazzotti}
\email[]{s.guazzotti14@imperial.ac.uk}
\author{Andreas Pusch}
\affiliation{Department of Physics, Imperial College London, London SW7 2AZ, UK }
\author{Doris E. Reiter}
\affiliation{Department of Physics, Imperial College London, London SW7 2AZ, UK }
\affiliation{Institut f\"ur Festk\"orpertheorie, Universit\"at M\"unster, Wilhelm-Klemm-Strasse 10, 48149 M\"unster, Germany}
\author{Ortwin Hess}
\email[]{o.hess@imperial.ac.uk}
\affiliation{Department of Physics, Imperial College London, London SW7 2AZ, UK }
\date{\today}

\newlength{\figsize}
\setlength{\figsize}{8.6cm}

\begin{abstract}
  Non-linear phenomena in optically excited semiconductor structures are of high interest. 
  We here develop a model capable of studying the dynamics of the photoexcited carriers, including Coulomb interaction on a Hartree-Fock level, on the same footing as the dynamics of the light field impinging on an arbitrary photonic structure.
  Applying this method to calculate the third harmonic generation in a semiconductor quantum well embedded in a Bragg mirror structure, we find that the power-law exponent of the intensity dependence of the third harmonic generation depends on the frequency of the exciting pulse.
  Off-resonant pulses follow the expected cubic dependence, while the exponent is smaller for resonant pulses due to saturation effects in the induced carrier density.
  Our study provides a detailed understanding of the carrier and light field dynamics during non-linear processes.
\end{abstract}


\maketitle

\section{Introduction}
Semiconductors exhibit a multitude of nonlinear optical responses for resonant as well as non-resonant excitation.\cite{Axt1998}
One of the most prominent nonlinear features is the generation of higher harmonics of the exciting frequency.
When the frequency of the incoming field is tripled one speaks of third harmonic generation (THG).
Such THG can be employed in spectroscopy and provides important insights into biological processes\cite{Yelin1999, Weigelin2016} or even for palaeontology.\cite{Chen2015}
In semiconductors, THG has, for example, been studied in coupled quantum wells,\cite{Sirtori1992,Heyman1994} quantum cascade structures,\cite{Mosely2004} quantum wires and dots,\cite{Sauvage1999,Wang2005} while it is also of interest in newly developed materials like graphene\cite{Hong2013} and atomically thin semiconductors.\cite{Wang2013}

In order to understand THG one requires a description of the optical fields and the material which is excited by them and generates the nonlinear interaction.
Here we focus on the photointeraction of semiconductor quantum wells (QW) with ultrashort light pulses.
To this end, we employ an auxiliary differential equation finite difference time domain (FDTD) approach to describe the dynamics of the light field along with the dynamics of the carriers in the QW.
This approach goes beyond rotating wave approximation and slowly-varying envelope approximation, allowing to treat fundamental and third harmonic on the same footing and describe photonic structures that vary on scales much smaller than the wavelength.
The combination of FDTD with density matrix models through auxiliary differential equations includes not only the effect of the field on the material but also self-consistently describes the effect of the material on the field.
This feature allows, for example, to describe propagation of SIT solitons in 1 and 2 dimensions\cite{Ziolkowski1995,Pusch2010,Pusch2011} and to study loss compensation and lasing dynamics in metamaterials\cite{Wuestner2010,Fang2010,Wuestner2011,Wuestner2012} or plasmonic stopped-light lasers.\cite{Pickering2014}
However, the few-level models employed in those studies can not describe the complicated behavior of an interacting electron gas excited in semiconductor QWs.\cite{Cundiff1994,Schuelzgen1999,Rossi2002}

On the other hand more complex wave-vector resolved semiconductor models have been developed that also consider Coulomb interaction between excited carriers within different levels of approximation\cite{Axt1998,Rossi2002,haug2004} or spatially resolved quantum kinetics calculations.\cite{Reiter2006,Reiter2007,Rosati2013}
Such models have been used to investigate various non linear effects such as the two-band Mollow triplet in thin GaAs films,\cite{Vu2004} the carrier-wave Rabi flopping in bulk GaAs\cite{Muecke2001} and THG from carbon nanotubes both in the perturbative and non-perturbative regime.\cite{Nemilentsau2006,Stanciu2002}
These approaches, however, do not include the self-consistent, spatially resolved resolution of electromagnetic fields.

Combining a spatially dependent full time-domain (FDTD) approach with a description of semiconductor QWs containing a wave-vector resolved, many-level density matrix description of the QW in a two-band approximation, has been pioneered in \citet{Boehringer2008,Boehringer2008a} to describe the spatio-temporal dynamics of semiconductor lasers and recently to describe lasing of semiconductor nanowires.\cite{Buschlinger2015}
Here, we extend the previous description by taking into account Coulomb interaction in Hartree-Fock approximation, which allows us to describe the excitonic nature of the QW absorption.

In this work, we are going to consider specifically the ultrashort pulse excitation of a QW embedded in a Bragg mirror structure typical for a semiconductor saturable absorber mirror (SESAM).
We obtain the carrier dynamics associated with excitation of the QW exciton and study the intensity dependence of THG in this QW.
We find that the power-law exponent of the intensity dependence of the THG strongly varies with excitation frequency.
For far off-resonant pulses the expected cubic behavior is found, while for pulses resonant with the exciton energy the exponent is reduced due to saturation effects.
Similar findings have been reported in theoretical and experimental studies on the excitation of carbon nanotubes with ultrashort laser pulses.\cite{Nemilentsau2006,Stanciu2002}

\section{Theory\label{sec:theory}}
The Hamiltonian describing the semiconductor structure is given by the three parts\cite{Rossi2002}
\begin{equation}
  \Hop = \Hop_{c} + \Hop_{cc} + \Hop_{c-l},
\end{equation} 
with the free carrier part $\Hop_{c}$, the carrier-carrier interaction $\Hop_{cc}$ and the carrier-light field interaction $\Hop_{c-l}$.
We assume a two-band structure with one conduction and one valence band, such that the free carrier Hamiltonian reads
\begin{equation}
  \Hop_{c} = \sum_{\kvec}\left[ \, \varepsilon^{e}_{\kvec} \eopd_{\kvec} \eop_{\kvec} + 
  \varepsilon^{h}_{\kvec} \hopd_{\kvec} \hop_{\kvec} \right].
\end{equation}
$\eopd_{\kvec}/\hopd_{\kvec}$ and $\eop_{\kvec}/\hop_{\kvec}$ are the electron/hole creation and annihilation operators with wave-vector $\kvec$ and $\varepsilon^{e/h}_\kvec$ are the corresponding energies.
We consider a QW, where the energy is quantized in the $z$-direction with a fixed $k_z$, while $\kvec$ always refers to the two-dimensional inplane wave vector $\kvec=(k_x,k_y,0)$.
The confinement along $z$ is included by applying the envelope function approximation,\cite{rossi2011} while the inplane bands are assumed to have parabolic dispersion.
The electron and hole energies are
\begin{subequations}
  \begin{align}
    \varepsilon^{e}_{k}&= \frac{\hbar^2 \kvec^2}{2 m_{e}} + \frac{\hbar^2 k^2_z}{2 m_{e}} + \varepsilon_{\text{gap}}, \\
    \varepsilon^{h}_{k}&= \frac{\hbar^2 \kvec^2}{2 m_{h}} + \frac{\hbar^2 k^2_z}{2 m_{h}},
  \end{align}
\end{subequations}
with the effective masses $m_{e/h}$ and the band gap $\varepsilon_{\text{gap}}$.  

The carrier-carrier interaction is given by the Coulomb potential
\begin{eqnarray} \nonumber
  \Hop_{cc} &=& \frac{1}{2} \sum_{\kvec, \kdashvec, \qvec} \left[ \Vkkd^{ee}
    \eopd_{\kvec+\qvec} \eopd_{\kdashvec-\qvec} \eop_{\kdashvec} \eop_{\kvec} \right. \\
    && + \Vkkd^{hh} \hopd_{\kvec+\qvec} \hopd_{\kdashvec-\qvec} \hop_{\kdashvec} \hop_{\kvec} \nonumber  \\
    && \left.- 2 \Vkkd^{eh} \eopd_{\kvec+\qvec} \hopd_{\kdashvec - \qvec}
  \hop_{\kdashvec} \eop_{\kvec}\right],
\end{eqnarray}
with the Coulomb matrix elements $\Vkkd^{ee/hh/eh}$ obtained by multiplying the ideal $2D$ Coulomb matrix elements by a band-dependent form-factor obtained from the envelope function approximation.
We consider the Plasmon-Pole\cite{haug2004} approximation to the screening of the Coulomb potential, where the inverse screening length is kept constant at the initial value, $\kappa_0$, as screening typically builds up on timescales longer than those considered here.\cite{ElSayed1994, Banyai1998}

We treat the carrier-light field interaction in dipole approximation resulting in
\begin{equation}
  \Hop_{c-l} = - \sum_{\kvec} d E(z_{\text{QW}};t) \left[ \eopd_{\kvec} \hopd_{-\kvec} + \hop_{-\kvec} \eop_{\kvec} \right],
\end{equation}
with dipole matrix element $d$ for the transition from valence to conduction band.
The classical light field $E (z_{\text{QW}};t)$ is assumed to be spatially constant over the region of the QW, denoted by the parametric dependence of the light field on $z_{\text{QW}}$.

To calculate the dynamics of the system we set up the equations of motion for the occupations $n^{e}_{\kvec} = \langle \eopd_{\kvec} \eop_{\kvec} \rangle$ and $n^{h}_{\kvec} = \langle \hopd_{\kvec} \hop_{\kvec} \rangle$ and the polarization $p_{\kvec} = \langle \hop_{-\kvec} \eop_{\kvec}  \rangle$ via the Heisenberg equation of motion
\begin{eqnarray}
  \label{eq:Bloch} \partial_{t} p_{\kvec} & = & - i \, \omega_{\kvec} p_{\kvec} - i \, \Omega_{\kvec}[n^e_{\kvec} + n^h_{\kvec} - 1] - \gamma_{p} p_{\kvec}, \nonumber \\
  \partial_{t} n^{e}_{\kvec} & = & i \, [\Omega_{\kvec} p^{*}_{\kvec} - \Omega^{*}_{\kvec} p_{\kvec}], \nonumber \\
  \partial_{t} n^{h}_{\kvec} & = & i \, [\Omega_{\kvec} p^{*}_{\kvec} - \Omega^{*}_{\kvec} p_{\kvec}].
\end{eqnarray}
Here $\omega_{\kvec}$ is the transition frequency, $\gamma_p$ is a phenomenological dephasing rate and $\Omega_\kvec$ is the Rabi frequency beyond rotating wave approximation, i.e., calculated with the time dependent electric field $\mathbf{E}\left(z_{\text{QW}};t\right)$.
Due to the homogeneity of the problem, we only take into account the $\kvec$-diagonal elements of the density matrix. The off-diagonal element are known to play a crucial role for spatially inhomogeneous problems.\cite{Reiter2006,Reiter2007,Rosati2013}

The equations of motion (Eq.~\ref{eq:Bloch}) already include Coulomb interaction under Hartree-Fock approximation, which is justified for ultra short time scales.
Within this approximation the interaction leads to a renormalization of the transition energies
\begin{equation}
  \label{eq:omegak}
  \hbar \omega_{\kvec} = \varepsilon^e_{k} +\varepsilon^h_{k}
  - \sum_{\kdashvec \neq \kvec} \left( \Vkkd^{ee} n^{e}_{\kdashvec} + \Vkkd^{hh} n^{h}_{\kdashvec}\right) + \delta E^{CH},
\end{equation}
with the Coulomb hole self-energy
\begin{equation}
  \delta E^{\text{CH}} = \sum_{q\neq 0} \left( V^s_{q} - V^b_{q} \right)
\end{equation}
and the bare (unscreened) Coulomb potential $V^{b}_{\left|\kvec-\kvec'\right|}$.
Also the light-matter coupling becomes renormalized due to the Coulomb interaction leading to the renormalized Rabi frequency
\begin{equation}
  \label{eq:rabik}
  \hbar\Omega_{\kvec} = d E(t) + \sum_{\kdashvec \neq \kvec} \Vkkd^{eh} p_{\kdashvec}.
\end{equation}

The integration of the equations of motion (Eq.~\ref{eq:Bloch}) is performed on a grid of $201$ $k$-points, homogeneously distributed between $k = 0~\text{m}^{-1}$ and $k = 15 / a_0$, where $a_0$ is the Bohr radius in the bulk material.
The integration algorithm is Runge-Kutta of order 4, where the Rabi frequency at the midpoints is obtained by interpolation of the electric field.\cite{Pusch2010}

In the simulation we are not only interested in the light field acting on the carriers in the QW, but also on the back-action on the field itself.
We model the dynamics of the electric field $\Efield$ in the whole structure as well as the ingoing and outgoing field through a one dimensional FDTD simulation.\cite{Taflove2005}
In the one-dimensional case, with the field propagating along $z$, Maxwell Equations can be reduced to
\begin{eqnarray}
  \label{eq:Maxwell1D}
  \frac{\partial H(z,t)}{\partial t} & = & \frac{\partial E(z,t)}{\partial z}, \\
  \frac{\partial E(z,t)}{\partial t} & = & \frac{1}{\epsilon_{b}(z)} \left[ \frac{\partial H(z,t)}{\partial z} - P(z,t)\right],
\end{eqnarray}
where $E\left(z,t\right)$ and $H\left(z,t\right)$ are the electric and magnetic fields and $\epsilon_{b}\left(z\right)$ is the background permittivity.
The dynamic material polarization $P\left(z,t\right) = P\left(z_{\text{QW}};t\right) \delta\left(z - z_{\text{QW}}\right)$ is zero everywhere but at the position of the QW.
$P\left(z_{\text{QW}};t\right)$ can be calculated from the microscopic polarizations as
\begin{equation}
  P\left(z_{\text{QW}};t\right) = 2 \sum_{\kvec} \text{Re}\left(p_{\kvec}\right) d.
  \label{eq:MacroscopicPolarisation}
\end{equation}
The spatial grid used to describe the system has a step of $dx = 10~\text{nm}$. Due to FDTD stability constraints this results in a time step of $dt \simeq 0.0333~\text{fs}$, which has been used for the simultaneous resolution of the semiconductor equations of motion and Maxwell Equations.
The injection of field inside the simulation domain is performed through the total field scattered field (TFSF) technique.\cite{Taflove2005}
The open boundaries of the system are simulated through perfectly matched layers (PML) boundary conditions.

\section{Results\label{sec:results}}
To test our model we will start by investigating a QW in a homogeneous background.
We will then study the field and semiconductor dynamics for a QW embedded in a multilayered structure.
In the simulation different passive materials are defined by a constant refractive index and different structures can be modeled by defining a space dependent refractive index profile.
The active medium we chose to investigate with our model is a $\text{In}_{0.2}\text{Ga}_{0.8}\text{As}/\text{GaAs}$ QW.
The parameters required for the simulation are listed in Table~\ref{tab:QW_parameters}.
The system is probed with pulses having a hyperbolic secant shape
\begin{equation}
  \label{eq:Pulse_Shape}
  E\left(t\right) = E_0 \cos\left(\omega_p t\right) \sech \left(t/\tau\right),
\end{equation}
with the pulse energy $\varepsilon_p = \hbar \omega_p$.
The full width half maximum (FWHM) of the pulse is $T \simeq 2 \tau \log\left(2+\sqrt{3}\right)$.

\subsection{Quantum well in homogeneous background}
\begin{figure}
  \includegraphics[width=\columnwidth]{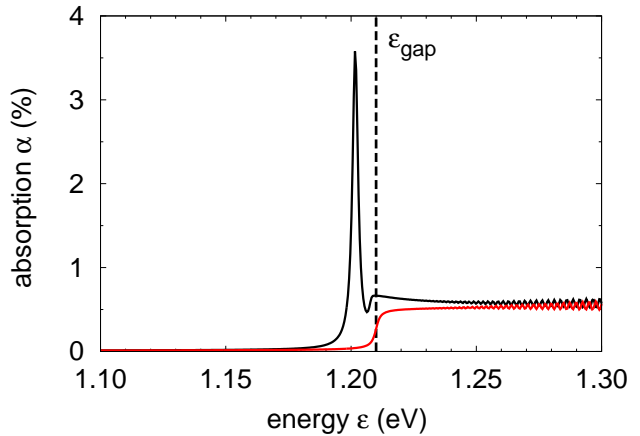}
  \caption{(color online) Linear absorption $\alpha$ as function of energy $\varepsilon$ of a single QW immersed in an infinitely extended background of GaAs with (black) and without (red) Coulomb interaction.\label{fig:OnlyQWAbsorption}}
\end{figure}
\begin{table}
  \begin{tabular}{lcc}
    \hline
    \hline
    effective electron mass &$m_e$& $0.06$ $m_0$\\
    effective hole mass &$m_h$& $0.33$ $m_0$\\
    QW width & $w$ & $10$ nm\\
    dipole matrix element & $d$ & $0.5$ nm\\
    band gap & $\varepsilon_{\text{gap}}$ & $1.21$ eV \\
    polarization dephasing & $\gamma_p$ & $2~\text{ps}^{-1}$ \\
    $n_{\text{In}_{0.2}\text{Ga}_{0.8}\text{As}}$ & $n_{\text{QW}}$ & $3.698$ \\
    $n_{\text{GaAs}}$ & $n_{\text{barrier}}$ & $3.5507$ \\
    \hline
    \hline
  \end{tabular}
  \caption{Parameters for the carrier dynamics in the QW, with the free electron mass $m_0$.\label{tab:QW_parameters}}
\end{table}
We start by analyzing a QW embedded in an homogeneous background of GaAs material filling the whole simulation domain, which is $2.01~\mu\text{m}$ long.
This allows us to focus on the properties of our model, namely the carrier dynamics in the semiconductor and the light field dynamics.
First we study the linear response of our system to a weak excitation.
We use a pulse with the central frequency close to the semiconductor band gap ($\varepsilon_{p} = \hbar \omega_p =1.2~\text{eV}$) and a FWHM of $15~\text{fs}$ to simulate a broad spectrum.
Our full time and spatial-domain description through an FDTD algorithm means that we have access to the full field dynamics in the simulation domain, including the back-action from active regions.
Through this, we see that for a QW in a homogeneous dielectric environment the fraction of reflected field is negligible and the incoming pulse is only transmitted and absorbed, i.e., $I_{\text{inc}} = I_{\text{trans}} + I_{\text{abs}}$ where $I_{\text{inc}}, I_{\text{trans}}\text{ and }I_{\text{abs}}$ are the incoming, transmitted and absorbed intensities, respectively,
\begin{equation}
  I\left(\varepsilon\right) = \frac{E^2\left(\varepsilon\right)}{c \mu_0},
\end{equation}
with the speed of light in vacuum $c$ and the vacuum permeability $\mu_0$.
With this we obtain the absorption spectrum as
\begin{equation}
  \alpha\left(\varepsilon\right) = 1 - \frac{I_{\text{trans}}\left(\varepsilon\right)}{I_{\text{inc}}\left(\varepsilon\right)}.
\end{equation}
The absorption spectrum for the QW in the case with and without Coulomb interaction is shown in Fig.~\ref{fig:OnlyQWAbsorption}.
The red line is the absorption of a non interacting system showing a step function smeared out around the band edge.
When Coulomb interaction is included in the simulation, we obtain the black line in Fig.~\ref{fig:OnlyQWAbsorption}, which shows a strong resonance below band edge.
Here the exciton, i.e., a bound state between electron and hole, is formed.
We calculate the binding energy of the exciton ground state as $\varepsilon_b = \varepsilon_{\text{gap}} - \varepsilon_X = 8~\text{meV} \simeq 2.16~\varepsilon_{\text{Ry}} $, where $\varepsilon_X$ is the position of the resonance in the absorption spectrum and $\varepsilon_{\text{Ry}} = 3.706~\text{meV}$ is the Rydberg energy of the exciton in the bulk material,\cite{haug2004} calculated with the parameters in Table~\ref{tab:QW_parameters}.
This energy is mainly determined by the QW thickness in relation to the effective Bohr radius of the material.\cite{Bastard1982}
The two limiting cases are an infinitely thin 2D QW, in which $\varepsilon_{b} = 4~\varepsilon_0$, and one that is thick enough to be considered a bulk material ($\varepsilon_{b} = \varepsilon_0$).
The width and height of the resonance are determined by the polarization dephasing rate, a faster dephasing (i.e. larger dephasing rate) results in broader less intense resonances, and the dipole matrix element.
The oscillations appearing for high energies in Fig.~\ref{fig:OnlyQWAbsorption} are due to the finite $k$-space resolution of the simulation.
The carrier dynamics sensitively depends on the excitation strength.
To quantify it we draw a comparison with a two level system.
Note that, when we neglect the Coulomb interaction the semiconductor model behaves as a set of two level systems with the same dipole matrix element.
The pulse area of a two level system with dipole matrix element $d$ is defined as
\begin{equation}
  \theta_{2} = \int_{-\infty}^{+\infty} \Omega\left(t\right) \dif t = \int_{-\infty}^{+\infty} \frac{E\left(t\right) d}{\hbar} \dif t =
  \frac{d E_0 \tau}{\hbar} \pi,
\end{equation}
where in the last step the integration was carried out for the hyperbolic secant shaped pulse (Eq.~\ref{eq:Pulse_Shape}).
A pulse of area $\theta = 2 \pi$ is defined as the pulse which increases the inversion of the system from the ground state, reaching maximum inversion at the pulse maximum, and then brings the system back to its ground state.

Due to the presence of Coulomb interaction in our model the Rabi frequency is renormalized and different among different states, i.e., dependent on $k$.
We thus extend the definition to our system by defining the area of the pulse with respect to the QW as
\begin{eqnarray}
  \theta_s &=& \frac{1}{N_k} \sum_{\kvec} \theta_{\kvec}, \\
  \theta_{\kvec} &=& \int \Omega_{\kvec}\left(t\right) \dif t \label{eq:thetak_1}\\
  &=& \theta_{2} + \sum_{\kvec'\neq\kvec} \Vkkd \int p_{\kvec'}\left(t\right) \dif t,\label{eq:thetak_2}
\end{eqnarray}
where we have used Eq.~\ref{eq:rabik} and $N_k$ is the number of states.
Nevertheless the area with respect to a two level system is still a good approximate measure of the pulse strength.
\begin{figure}
  \includegraphics[width=\columnwidth]{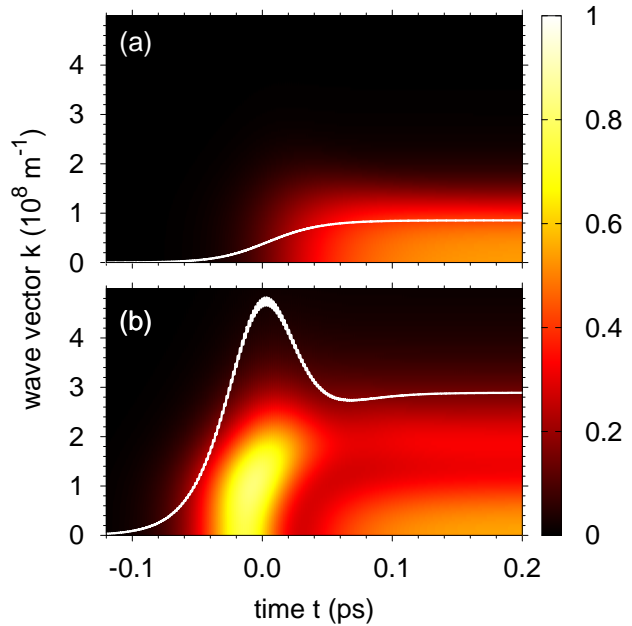}
  \caption{(color online) Time evolution of the occupation of electron states (colormap) and of the total electronic density (white line) for (a) a weak pulse with peak intensity $I_0\simeq 1.1\times10^{12} \frac{\text{W}}{\text{m}^2}$ and (b) a strong pulse with peak intensity $I_0\simeq 2.7\times10^{13} \frac{\text{W}}{\text{m}^2}$.\label{fig:OnlyQWDensityKRes}}
\end{figure}

Next we investigate the dynamic behavior of the system under excitation from stronger pulses, i.e., in the non-linear regime.
Figure~\ref{fig:OnlyQWDensityKRes}(a) shows the electron dynamics for a weak excitation with pulse area $\theta_{2} \simeq 0.4\pi$.
The exciting pulse is resonant with the exciton and has a FWHM of $100~\text{fs}$ and a peak intensity of $I_0 \simeq 1.1\times10^{12}~\frac{\text{W}}{\text{m}^2}$.
We use the time at which the pulse maximum reaches the QW as the zero of our time scale.
The white overlay in Fig.~\ref{fig:OnlyQWDensityKRes} is the total density of carriers in the system, which reads
\begin{equation}
  N = \frac{1}{2\pi} \int g\left(k\right) n^e_k \dif k,
\end{equation}
where $g\left(k\right) \dif k$ is the number of states between $k$ and $k + \dif k$.
From the monotonous increase of density through the pulse action and the absence of Rabi oscillations, we deduce that the system behaves like a two level system excited by a pulse with area $\theta_{s} < \pi$.
A more refined picture is given by the color map in Fig.~\ref{fig:OnlyQWDensityKRes}(a) which shows the occupation of the electronic states as a function of time and wave vector $k$.
The population is located around the minimum of the conduction band, $k = 0$, and its distribution in $k$ space does not significantly change shape while the pulse traverses the material.

The electron dynamics represented in Fig.~\ref{fig:OnlyQWDensityKRes}(b) is obtained with a stronger excitation, $25$-fold increase in the peak intensity, i.e., $\theta_{2} \simeq 2\pi$.
Just by looking at the total density one concludes that the semiconductor as a whole behaves like a two level system, i.e., it shows Rabi oscillations with an effective area of $\theta_{s} \simeq 2\pi$.
A more complete picture comes from looking at state occupations.
First of all we see that the population extends itself further from the band minimum with respect to the weaker excitation, the reason being that the spectrum of the more intense pulse has stronger high energy components.
We also observe that every state behaves as a two level system excited by a pulse with a different effective area.
This is the consequence of the Rabi frequency becoming strongly $k-$dependent and we attribute this to two main reasons.
Firstly a variation of the Rabi frequency between different states is due to the introduction of Coulomb interaction as is exemplified by Eq.~\ref{eq:rabik}.
Secondly, Rabi oscillations occur at the generalized Rabi frequency and thus depend on the detuning between the central wavelength of the pulse and the optical transition.
This is of particular importance for our system because of the underlying parabolic dispersion, meaning that the detuning grows approximately as $k^2$.
\begin{figure}
  \includegraphics[width=\columnwidth]{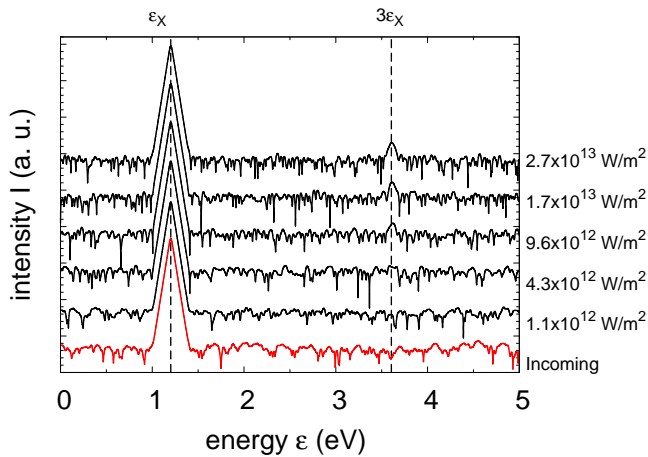}
  \caption{(color online) Spectrum of the transmitted pulse in the non-linear regime on semilogarithmic scale. The different lines are artificially displaced and are obtained by changing the peak intensity of the incident pulse. From bottom to top, the peak intensities of the black spectra are $\left(1.1, 4.3, 9.6, 17, 27\right)\times10^{12} \frac{\text{W}}{\text{m}^2}$. The red line is the spectrum of the incoming pulse for a peak intensity of $1.1\times10^{12} \frac{\text{W}}{\text{m}^2}$.\label{fig:OnlyQWSpectra}}
\end{figure}

The advantage of our model is that it includes a feedback from the semiconductor to the field, which can give rise to non-linear effects in the transmitted field.
To investigate this we start with a weak pulse resonant with the exciton (with peak intensity $I_0 \simeq 1.1\times10^{12}~\frac{\text{W}}{\text{m}^2}$ and $100~\text{fs}$ FWHM) and analyze the transmitted field.
The bottom black line in Fig.~\ref{fig:OnlyQWSpectra} shows the square of the field spectrum, which is proportional to the intensity, as a function of photon energy $\varepsilon$.
This spectrum shows a single peak located around $1.2~\text{eV}$ and corresponding to the spectrum of the incoming pulse (red line in Fig.~\ref{fig:OnlyQWSpectra}), resonant with the exciton, $\varepsilon_0 = \varepsilon_X = 1.202~\text{eV}$.
\begin{figure}
  \includegraphics[width=\columnwidth]{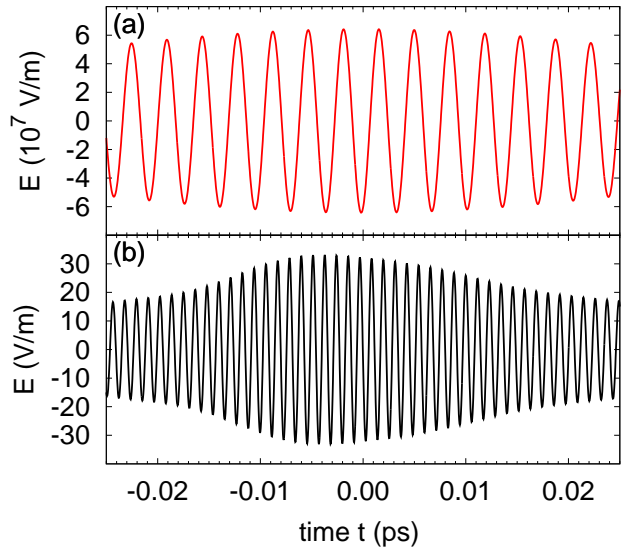}
  \caption{(color online) (a) Total field transmitted through the QW.
    (b) Filtered third harmonic component of the transmitted field.\label{fig:OnlyQWTHGPulse}}
  \end{figure}
Next we increase the intensity of the exciting pulse and calculate the spectrum of the transmitted field, to get the series of lines in Fig.~\ref{fig:OnlyQWSpectra}.
Going from bottom to top (excluding the red spectrum of the incoming pulse), the peak intensity of the exciting pulse are $\left(1.1, 4.3, 9.6, 17, 27\right)\times10^{12} \frac{\text{W}}{\text{m}^2}$.
As the pulse intensity increases, a second peak centered at $\varepsilon_3 = 3 \varepsilon_0$ emerges over the background, signaling the presence of THG in the model.
For the range of excitation intensities that we explored, the third harmonic pulse is several orders of magnitude smaller than the incoming pulse and thus its presence is not appreciable in the transmitted field, as shown in Fig.~\ref{fig:OnlyQWTHGPulse}(a).
It is possible to obtain the third harmonic field in time domain by applying a bandpass filter around the energy $\varepsilon_3$.
The resulting pulse is shown in Fig.~\ref{fig:OnlyQWTHGPulse}(b).

The results in Figures~\ref{fig:OnlyQWSpectra} and \ref{fig:OnlyQWTHGPulse}(b) show that the our model is suitable for the study of optical non-linearities in semiconductors.

\subsection{Quantum well inside SESAM}
\begin{figure}
  \includegraphics[width=\columnwidth]{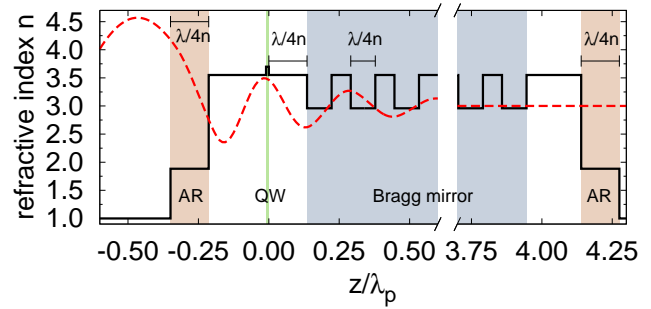}
  \caption{(color online) Refractive index profile of the SESAM structure (black solid line) and field profile (red dashed line).\label{fig:structure}}
\end{figure}
\begin{table}
  \begin{tabular}{lcc}
    \hline
    \hline
    Material & $n$ & $\epsilon_b$ \\
    \hline
    SiN & $1.884$ & $3.551$\\
    $\text{In}_{0.2}\text{Ga}_{0.8}$As & $3.698$ & $13.675$ \\
    GaAs & $3.551$ & $12.610$ \\
    AlAs & $2.959$ & $8.756$ \\
    \hline
    \hline
  \end{tabular}
  \caption{Refractive index and background permittivity of all the materials included in the simulation.\label{tab:refractive_indices}}
\end{table}
After studying the electron dynamics of an isolated QW in a homogeneous background we proceed by introducing a more realistic optical environment.
For this we choose a SESAM, which is a well established structure for ultra short pulse generation.\cite{Keller2006}
The whole structure is included in our simulations as a spatially varying background permittivity $\epsilon_b\left(z\right)$, as shown in Fig.~\ref{fig:structure}, and is surrounded by $2~\mu \text{m}$ of air on each side.
Table~\ref{tab:refractive_indices} contains the refractive index and background permittivity of all materials included in the structure.

\begin{figure}
  \includegraphics[width=\columnwidth]{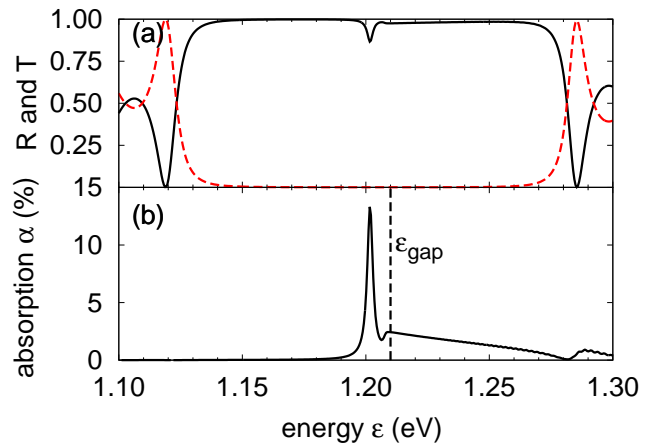}
  \caption{(color online) (a) Reflection R (black solid line) and transmission T (red dashed line) coefficients of the SESAM structure.
    (b) Linear absorption spectrum. The vertical dashed line indicates the gap energy.\label{fig:QW_absorption}}
\end{figure}

In order for the structure to be effective, most of the layer thickness need to be proportional to the central wavelength of the incoming pulse, $\lambda$.
We coated both ends of the structure with a SiN layer of optical length $\lambda/4$ which minimizes reflection of the pulse coming from air.
This allows for a more efficient in-coupling of the light.
The mirror is composed by a set of alternating GaAs and AlAs layers, each with an optical length of $\lambda/4$.
This basic two-layered module is repeated $25$ times in order to achieve a very high reflectivity around $\lambda$, as shown in Fig.~\ref{fig:QW_absorption}(a), where the pulse is resonant with the exciton energy $\varepsilon_X$.
Due to the presence of the mirror, a standing wave is created inside the GaAs layer between the mirror itself and the anti-reflective layer.
Such an interference pattern has zeros at even integer multiples of $\lambda/4n$ and maxima at odd ones.
The QW is then located in such a maximum, in order to take advantage of the field enhancement provided by this interference pattern, and is thus at an optical distance $\lambda/4$ from the start of the mirror.
Two further layers of arbitrary size are used to isolate the anti-reflection coatings from the mirror on one side and the QW on the other.

We start again by analyzing the linear regime where we use a pulse with $\varepsilon_p = 1.2~\text{eV}$ and a FWHM of  $15~\text{fs}$.
From the reflected and transmitted field we obtain the spectra of transmittance ($T$) and reflectance ($R$) of the structure, shown in Fig.~\ref{fig:QW_absorption}(a).
We see that the mirror used in the simulation reflects almost perfectly in a broad spectral region around the central wavelength of the pulse.
We further calculate the absorption of the structure as $\alpha = 1 - R - T$, shown in Fig.~\ref{fig:QW_absorption}(b).
The absorption is mostly determined by the QW and thus we find a similar behavior as in Fig.~\ref{fig:OnlyQWAbsorption} with a resonance at the exciton energy.
The resonance height is increased by a factor of $3.7$ in comparison with the isolated QW, as a consequence of the almost four-fold enhancement in intensity introduced by the structure, while the in-band absorption is now decreasing the further we go from the band edge.
The difference is due to the mirror which is optimized for the central wavelength of the pulse and whose reflectance decreases with the distance from the exciton.
\begin{figure}
  \includegraphics[width=\columnwidth]{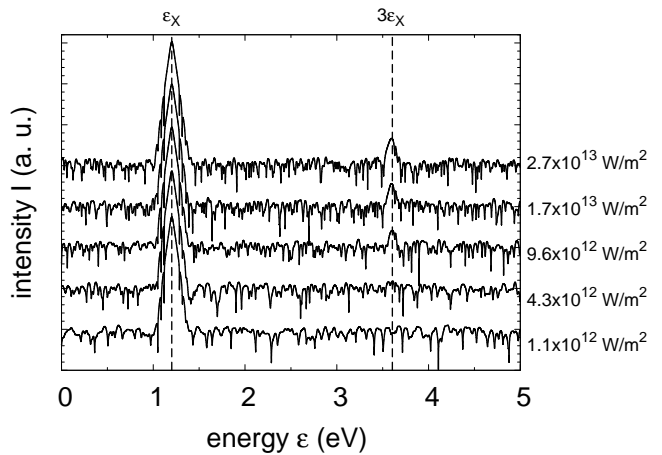}
  \caption{(color online) Spectra of the field reflected from the SESAM structure on semilogarithmic scale. The lines correspond to increasing peak intensity of the incoming pulse, from bottom to top $\left(1.1, 4.3, 9.6, 17, 27\right)\times10^{12} \frac{\text{W}}{\text{m}^2}$\label{fig:THG_spectra}}
\end{figure}

Next, we shine a set of subsequently stronger pulses on the SESAM structure to investigate the non-linear regime.
The pulses are resonant with the exciton ($\varepsilon_p = \varepsilon_X$) and have a FWHM of $100~\text{fs}$.
Figure~\ref{fig:THG_spectra} shows the spectra of the reflected intensity for the same peak intensities used in Fig.~\ref{fig:OnlyQWSpectra}, namely $\left(1.1, 4.3, 9.6, 17, 27\right)\times10^{12} \frac{\text{W}}{\text{m}^2}$.
Similarly to what happened for the isolated QW, increasing the intensity of the exciting pulse brings a second spectral peak above the background.
This is located at three times the pulse energy and is due to THG in the semiconductor layer of the structure.
By comparing the spectra with Fig.~\ref{fig:OnlyQWSpectra} we see that the third harmonic is more intense in the SESAM structure than it is for an isolated QW excited with the same pulse.
This is due to the structure enhancing the field at the QW position.
The oscillations appearing in Fig.~\ref{fig:THG_spectra}, particularly evident in the fundamental peak, are due to the Fabry-P\'{e}rot resonance associated with the whole structure.
\begin{figure}
  \includegraphics[width=\columnwidth]{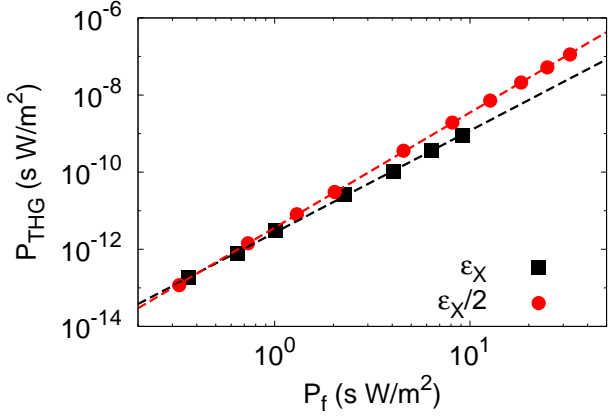}
  \caption{(color online) THG spectrally integrated intensity $P_{\text{THG}}$ as a function of the spectrally integrated intensity $P_{f}$ of the incoming pulse on a log-log scale for $\varepsilon_p = \varepsilon_X$ (black squares) and $\varepsilon_p = \varepsilon_X/2$ (red circles). The dashed lines are the best fit of the data according to Eq.~\ref{eq:intensity_dependence_linear}.\label{fig:THG_intensity}}
\end{figure}

A more quantitative analysis of the intensity is given in Fig.~\ref{fig:THG_intensity}, where we plot the integrated intensity of the THG as function of the incoming pulse integrated intensity.
The two sets of data correspond to different excitation energies, where one is obtained with pulses resonant with the exciton $\varepsilon_p = \varepsilon_X$ (squares) and one with off-resonant pulses with $\varepsilon_p = \varepsilon_X/2$ (circles).
The spectrally integrated intensity of the fundamental and its third harmonic are defined as the integral of the intensity over the corresponding spectral peak,
\begin{subequations}
  \begin{align}
    P_{f} &= \frac{1}{\hbar}\int_{f} I_r\left(\varepsilon\right) \dif \varepsilon, \\
    P_{\text{THG}} &= \frac{1}{\hbar}\int_{\text{THG}} I_r\left(\varepsilon\right) \dif \varepsilon,
  \end{align}
\end{subequations}
where the integration is carried out over the width of the highest intensity peak.
We find that the intensity of the THG has a power law behavior as function of the incoming field, $P_f$, as 
\begin{equation}
  \label{eq:intensity_dependence}
  P_{\text{THG}} \propto P^\delta_f.
\end{equation}
In the log-log plot this is seen as a linear curve, where we can get the value of the exponent by performing a linear fit of the logarithm of the data according to
\begin{equation}
  \label{eq:intensity_dependence_linear}
  \log\left(P_{\text{THG}}\right) = A + \delta \log\left(P_f\right),
\end{equation}
which gives the dashed lines in Fig.~\ref{fig:THG_intensity}.
The values obtained for the exponents (slope of the lines) are $\delta = 2.987 \pm 0.009$ for the off-resonant configuration and $\delta = 2.65 \pm 0.03$ for the pulses resonant with the exciton energy.
\begin{figure}
  \includegraphics[width=\columnwidth]{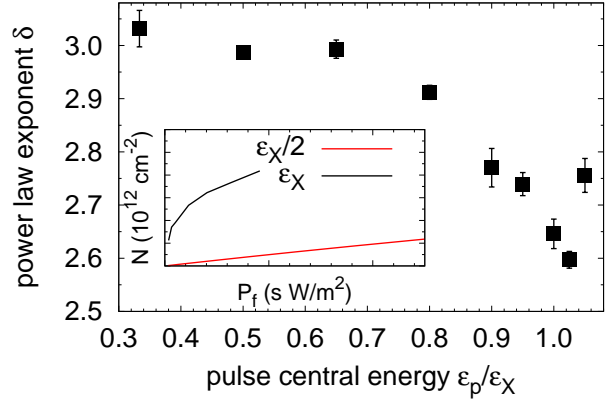}
  \caption{(color online) Power law exponent $\delta$, obtained by fitting Eq.~\ref{eq:intensity_dependence_linear}, as function of the pulse energy $\varepsilon_p$.
  The inset shows the residual density as a function of the spectrally integrated intensity for resonant and off-resonant excitations.\label{fig:THG_pulse}}
\end{figure}

Figure~\ref{fig:THG_pulse} shows the value of the exponent $\delta$ as a function of the central wavelength of the pulse.
Similarly to Fig.~\ref{fig:THG_intensity} we have performed a linear fit of Eq.~\ref{eq:intensity_dependence_linear} to different sets of data.
We see that when the excitation is enough off-resonant, i.e., the pulse energy lies in the band gap of the semiconductor, the value of $\delta$ approaches an asymptotic value of $\delta \sim 3$ which is consistent with a phenomenological description in terms of the non-linear susceptibility of third order, $\chi^{\left(3\right)}$.\cite{Shen1984,Schafer2002}
Conversely, as the excitation gets closer to be resonant we observe a decrease in $\delta$ down to a value of about $2.6$.
A similar non cubic dependence has been observed by \citet{Haase2000} while performing four wave mixing experiments on ZnSe QWs with the central laser energy close to resonance with the exciton.
We want to stress that we are able to uncover this unusual behavior only due to a self-consistent combination of the Bloch equations with a description of the light field using a full time-domain (FDTD) code with spatial resolution on sub-wavelength scales.

To further investigate the origin of this subcubic dependence, we analyzed the density of carriers generated by pulses with different detuning from the excitonic resonance.
The inset of Fig.~\ref{fig:THG_pulse} shows the density remaining in the semiconductor after the pulse has left the simulation domain as a function of the spectrally integrated intensity of the exciting pulse.
We see that the amount of population generated in the QW is significantly higher under resonant excitation, even for the lowest intensity generating a THG signal, than the population for the off-resonant excitation via more intense pulses.
Also the amount of population excited in the QW rises linearly with the pulse intensity for off-resonant excitation, while it shows a marked saturation behavior under resonant excitation.
In order to test whether the power law changes for smaller intensities our numerical simulation allows us to repeat the same analysis analyzing the transmitted field which has a lower level of background noise.
We observe a crossover from $\delta = 3$ for low intensities, to $\delta \neq 3$ at higher intensities.
We find that the minimum intensity for which an exponent different from 3 is obtained is $10^{13}~\text{W/m}^2$.
We have also checked that the crossover position is independent of the dephasing time.
Because of this and the correlation with the density of carriers in the QW, we attribute the change of the power law exponent $\delta$ to the presence of optically excited carriers and to the saturation of the total density in the semiconductor.

\section{Conclusions}
In summary, we have studied the emergence of third harmonic signals in semiconductor quantum wells (QW), photo-excited by intense femtosecond optical pusles.
For this, we have introduced a general model combining a full time and space dependent finite-difference time-domain (FDTD) description of the light field, i.e., a discretization of Maxwell's equations without the inherent limitations of the slowly-varying envelope approximation, with a wave-vector resolved many level and many-body density matrix approach for the charge carrier dynamics.
For a QW embedded in a homogeneous background we studied the interplay of light field dynamics and carrier dynamics, demonstrating the emergence of non-linear optical effects such as third harmonic generation (THG).
We further analyzed the intensity dependence of the generated non-linear response for a QW embedded in a many-layer semiconductor saturable absorber mirror (SESAM) structure and show that the intensity dependence of the THG signal strongly varies with excitation frequency.
For an excitation well below the band gap of the QW, we found that the intensity of the THG signal follows a cubic dependence on the intensity of the exciting pulse.
This is in direct agreement with a description based on an expansion in powers of the field with non-linear susceptibilities as constant coefficients.
For a resonant excitation at the excitonic frequency, however, the intensity dependence still follows a power-law, now with an exponent that is reduced to $2.6$, clearly deviating from the cubic behavior.
Although a non-cubic dependence can also be obtained with a more phenomenological approach of an intensity dependent $\chi^{\left(3\right)}$ coefficient,\cite{Al-Naib2014,Cheng2015} this can only be fit to existing data rather than emerge from a more fundamental model.
The simultaneous description of the light field and carrier dynamics not only allows for a deeper understanding of non-linear optical effects but is also readily expandable to other 2-dimensional semiconductor systems such as graphene, transition metal dichalcogenides\cite{Wang2013} or more complex structures like combined plasmonic-semiconductor structures.\cite{Feng2014}

\section*{Acknowledgments}
DER gratefully acknowledges support from the German Academic Exchange Service (DAAD) within the P.R.I.M.E. programme. 
This study was partially support by the Air Force Office of Scientific Research (AFOSR), and the European Office of Aerospace Research and Development (EOARD) is also acknowledged. 


%
\end{document}